\documentclass[%
aps,pre,amsmath,amssymb,showpacs,
twocolumn,
longbibliography,
showkeys
]{revtex4-2}

\usepackage{lineno}
\newcommand*\patchAmsMathEnvironmentForLineno[1]{
  \expandafter\let\csname old#1\expandafter\endcsname\csname #1\endcsname
  \expandafter\let\csname oldend#1\expandafter\endcsname\csname end#1\endcsname
  \renewenvironment{#1}
     {\linenomath\csname old#1\endcsname}
     {\csname oldend#1\endcsname\endlinenomath}}
\newcommand*\patchBothAmsMathEnvironmentsForLineno[1]{
  \patchAmsMathEnvironmentForLineno{#1}
  \patchAmsMathEnvironmentForLineno{#1*}}
\AtBeginDocument{
\patchBothAmsMathEnvironmentsForLineno{equation}
\patchBothAmsMathEnvironmentsForLineno{align}
\patchBothAmsMathEnvironmentsForLineno{flalign}
\patchBothAmsMathEnvironmentsForLineno{alignat}
\patchBothAmsMathEnvironmentsForLineno{gather}
\patchBothAmsMathEnvironmentsForLineno{multline}
}

\usepackage{graphicx}
\usepackage{dcolumn}
\usepackage{bm}
\usepackage{subfigure}
\newif\iffigure
\figurefalse
\figuretrue

\allowdisplaybreaks[3]

\def\g1{\gamma_1}

\def\rnd#1#2{\frac{\partial #1}{\partial #2}}

\def\et{\bm{e}_{\theta}}
\def\er{\bm{e}_{r}}
\def\Re{\mathit{Re}}

\def\trac{\mathrm{tracer}}
\def\targ{\mathrm{target}}
\def\mi#1{$^{-#1}$}

\begin{document}

\preprint{\today, ver.~\number\time}

\title{
Thermo-osmotic slip flows around a thermophoretic microparticle\\ 
characterized by optical trapping of tracers}

\author{Tetsuro Tsuji}
\email{tsuji.tetsuro.7x@kyoto-u.ac.jp}
\author{Satoshi Mei}%
\author{Satoshi Taguchi}%
\affiliation{%
Graduate School of Informatics, Kyoto University, Kyoto 606-8501, Japan
}%


\date{\today}
\begin{abstract}
Thermo-osmotic flow around a microparticle in a liquid is characterized by observing and analyzing the distribution of tiny particles, i.e., tracers, near the microparticle's surface. First, an optical trapping laser is used to localize the tracer motion along a circular path near the circumference of the microparticle. Then, upon creating an overall temperature gradient in the liquid, the tracers on the circular path, originally uniformly distributed, gather towards the hotter side of the microparticle, indicating a flow along the particle toward the hot. Analyzing the tracer distribution further, it is found that (i) the flow magnitude decreases with the distance from the surface, and (ii) changing the surface property of the microparticle results in a change in the flow magnitude. These show that the observed flow is a thermally induced slip flow along the microparticle's surface. Then, assuming a simple slip boundary condition for a fluid equation, we evaluate the magnitude of the slip coefficient based on two experimental data: (i) the thermophoretic velocity of the microparticle and (ii) the thermo-osmotic flow around the microparticle. The results of the two approaches are in quantitative agreement. They are also compared with those of theoretical models for a slip flow in existing studies.

\end{abstract}

\maketitle


When a fluid is in contact with a solid surface with inhomogeneous temperature distribution, thermally-induced slip flows manifest. These flows are often called thermo-osmosis or thermo-osmotic flows in the literature \cite{Bregulla2016,Fu2017,Proesmans2019,Anzini2019,Anzini2022}, being considered novel tools for micro/nanoscale material transport \cite{Mousavi2019,Trivedi2022,Xu2023} and manipulation \cite{Lou2018,Fraenzl2022}. 
 Thermo-osmosis has multiple physical origins such as an excess enthalpy variation \cite{Derjaguin1987,Anderson1989}, thermo-electric \cite{Fayolle2008,Rasuli2008,Wuerger2008}, and Marangoni effects \cite{Ruckenstein1981,Wuerger2007,Kim2022}. 

A growing interest in thermo-osmosis originates from recent development on photothermal effects and thermoplasmonics \cite{Baffou2017,Baffou2020,Kotsifaki2022}, where fluids and/or solid surfaces can be heated locally in nanoscale. 
These localized heat sources lead to a steep temperature gradient, inducing thermo-osmotic flows of the detectable order of magnitude, e.g., a few tens of \textmu m~s\mi{1} \cite{Fraenzl2022}.
Nonetheless, due to multi-physics nature and difficulties in experimental evaluation, comparisons between theory and experiment are considered challenging. 
To tackle this situation, the present paper develops a new tool for the experimental evaluation of thermo-osmotic slip flows, and apply it to the characterization of the flows around microparticles. 

Before explaining the developed tools, let us describe why we are interested in the flows around particles. Thermo-osmosis around particles is particularly important among its applications, because it is one of the driving mechanisms of thermophoresis of tiny particles \cite{Piazza2008,Piazza2008a,Wuerger2010,Tsuji2017} and active matters \cite{Jiang2010,Bickel2013,Auschra2021b,Paul2022a}. 
Thermophoresis has been widely explored for separating a mixture \cite{Vigolo2010b,Maeda2011,Tsuji2018b,Sanjuan2022}, molecular characterization in bioscience \cite{Wienken2010,Seidel2012,Liu2019}, and opto-thermal manipulation of nanomaterial \cite{Lin2018a,Setoura2019} and biomaterial \cite{Lin2017a}. 
Nevertheless, to the extent of authors' knowledge, the profiles of thermo-osmotic flows around thermophoretic particles have not been evaluated, and thus, the contribution of the flow to thermophoresis remains unclear.  
The motivation of the present paper is to present the direct connection between the thermophoresis of a particle and the thermally-induced flow around it. 
In Refs.~\cite{Weinert2008} and \cite{Jiang2010}, the thermally-induced slip flows around a micro- and Janus particle were observed, respectively, but the flow profile was not quantified. Thermoosmotic flows on the flat surface were observed in \cite{Bregulla2016,Fraenzl2022} and that around a Janus particle in \cite{Bregulla2019}. However, because the thermally-slip flows occur near the surface, tracers for flow visualization were scarce, preventing a systematic flow characterization.   

\begin{figure*}[tb]
\begin{center}
\iffigure
\includegraphics[width=\textwidth]{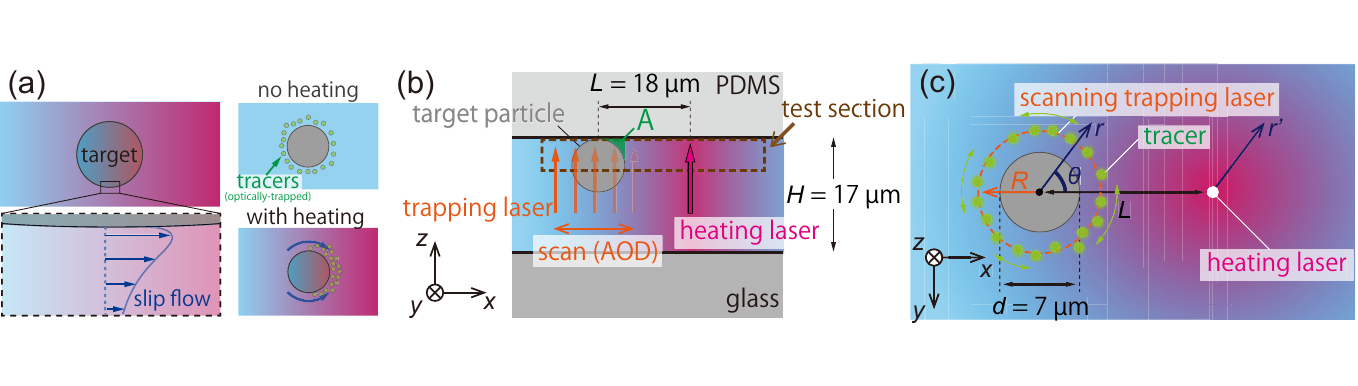}
\fi
\caption{
(a) Concept of a proposed scheme for the investigation of thermally-induced flow around a microparticle. Tracers are optically trapped around the microparticle, and play the role of flow indicators upon laser heating of a surrounding fluid. (b,c) Schematic of experiments:  
(b) A side view and (c) a bottom view (i.e., a camera view) of a test section. 
(b) A target particle with a diameter of $7$~\textmu m is immobilized at the upper surface of a microchannel with a height of $17$~\textmu m. A heating laser is irradiated at $L=18$~\textmu m away from the center of the target. (c) A scanning trapping laser on a circular path with a radius of $R$ confine fluorescent tracers around the target microparticle. 
}\label{fig:schematic}
\end{center}
\end{figure*}

Now we explain the proposed tool in this paper. To overcome the experimental difficulty above, we use optical trapping (or optical tweezer) \cite{Ashkin1971,Ashkin1986}. In recent years, optical trapping has been a powerful tool for the control of tiny objects \cite{Tsuji2022,Volpe2023}. 
Here, using the optical tweezer, we trap tracers as flow indicators (Fig.~\ref{fig:schematic}(a)). 
By confining the tracers optically in a test section, which is a circular path around a target microparticle, we facilitate the visualization of thermally-induced flows near the particle surface immersed in a water solution. 
In particular, we focus on the spatial dependence of the flow, the surface characteristics of the microparticle, and the connection between thermo-osmosis and thermophoresis.






\section{\label{sec:expcs}Experimental method}

We briefly explain the present experimental method, leaving details in Appendix~\ref{sec:exp_detail}.
A target polystyrene particle with a diameter of $d=7$~\textmu m is immobilized at the upper surface of a microchannel with a height of $17$~\textmu m, as shown in Fig.~\ref{fig:schematic}(b). 
The immobilization is necessary to focus on a flow around the target particle without its translocation. 

Then, a focused trapping laser with a wavelength of 1064~nm is irradiated to the upper surface of the microchannel. 
The trapping laser is scanned circularly around the target particle so that polystyrene fluorescent tracers with a diameter of $500$~nm are optically trapped around the target particle, as shown in Fig.~\ref{fig:schematic}(c).
A scanning circular path has a radius $R$, which is an experimental parameter and chosen as $R=4.4$, $5.4$, $6.5$, or $7.5$~\textmu m. 
We use a polar coordinate system with a radial position $r$ and an angle $\theta$ measured from a line connecting the center of the target particle and a heating spot (Fig.~\ref{fig:schematic}(c)).
The optical trapping is loose on the circular path, on which the tracers undergo the Brownian motion in the $\theta$ direction (Fig.~\ref{fig:schematic}(c)). 

After a while, many tracers are held on the circular path so that the spatial distribution of tracers become almost uniform with respect to $\theta$, as shown in Fig.~\ref{fig:schematic}(a). 
Then, at time $t=0$~s, a focused heating laser with a wavelength of $1480$~nm and with a power $P$ is irradiated to the upper surface of the microchannel at a position distant from the target particle (Fig.~\ref{fig:schematic}(b,c)). 
The temperature gradient of the fluid is kept by a photothermal conversion: the absorption of a near-infrared laser to water \cite{Duhr2006,Cordero2009,Riviere2016,Tsuji2018a}. Such optical heating is necessary to achieve a steep temperature gradient for the generation of the thermo-osmotic flow with a detectable magnitude. 

The distance between the target particle and the focus of the heating laser is set to $L=18.0\pm0.1$~\textmu m, and the distance from the heating laser is denoted by $r^\prime$ (Fig.~\ref{fig:schematic}(c)). 
The results of the temperature measurement for the heating laser power $P=13.5$, $30.2$, $46.7$, and $78.7$~mW are summarized in the Appendix~\ref{sec:temp}; the magnitude of the temperature gradient is $\sim 1$~K~\textmu m\mi{1}. 
We observe the change of the tracer distribution on the circular path in the $x\,y$ plane (Fig.~\ref{fig:schematic}(c)) after the onset of laser heating.

\section{\label{sec:model}Model}
In this section, we present a simple model that describes the physical situation of the experiment. For simplicity, we consider a steady state, which is realized after a few tens of seconds of the heating laser irradiation. Furthermore, we neglect the presence of microchannel wall and consider the temperature variation directed only to the $x$ direction.

\subsection{Fluids}

The temperature of the fluid and that inside the target particle are denoted by $T_f$ and $T_s$, respectively. 
Since the thermal P\'eclet number $\mathit{Pe}_T$ is small, i.e., $\mathit{Pe}_T\ll1$, we can neglect the convection term; the temperature fields $T_f$ and $T_s$ obey the steady heat conduction equations. 
For instance, when the reference length is $10$~\textmu m, the reference speed is $1$~\textmu m, and the thermal diffusivity is $1.4\times 10^{-7}$~m$^2$~s\mi{1}, we have $\mathit{Pe}_T = 7\times 10^{-5}$. 

The condition for the temperature imposed at infinity is $T_f\to T_0 + \alpha x$ $(r\equiv |\bm{x}| \to\infty)$ with a constant temperature gradient $\alpha(>0)$. 
Note that the $x$ direction corresponds to $\theta=0$ in Fig.~\ref{fig:schematic}(c), and $T_0$ is a reference temperature at the center of the target particle.
At the surface of the target particle $r=a(=d/2)$, the temperature and the heat flux in the normal direction to the sphere surface are continuous. Then, we readily obtain
\begin{subequations}\label{eq:temperature}
\begin{align}
&\frac{T_f -T_0}{\alpha a} = \left(1+\frac{a^3}{r^3}\xi_f\right)\frac{r}{a}\cos\theta, \quad 
\xi_f\equiv\frac{\kappa_f-\kappa_s}{2\kappa_f+\kappa_s}, \label{eq:temperature_fluid}\\
&\frac{T_s -T_0}{\alpha a} = \xi_s \frac{r}{a}\cos\theta, \quad 
\xi_s \equiv \frac{3 \kappa_f}{2\kappa_f+\kappa_s}(=\xi_f+1),
\end{align}
\end{subequations}
where $\alpha a$ is a characteristic temperature difference over a distance $a$; $\kappa_f$ and $\kappa_s$ are the thermal conductivity of the fluid and the target particle, respectively; $\xi_f$ and $\xi_s$ are non-dimensional parameters related with the thermal conductivity. 
In this paper, the temperature dependence of thermal conductivity is not considered, i.e., the constants $\kappa_f = 0.6$~W~m\mi{1}~K\mi{1}, $\kappa_s = 0.2$~W~m\mi{1}~K\mi{1}, $\xi_f\approx0.286$, and $\xi_s\approx1.286$ are used. 

The temperature field \eqref{eq:temperature} induces a fluid flow through a slip boundary condition, that is, the flow velocity vector $\bm{u}$ satisfies, on the sphere surface,
\begin{equation}
\begin{cases}
u_r\equiv\bm{u}\cdot\er=0, \\ u_\theta \equiv \bm{u}\cdot \et = - K \et \cdot \nabla T_f,  
\end{cases}
\quad (r=a),
\label{eq:utheta_bc}
\end{equation}
where $\er$ and $\et$ are the unit vectors in the $r$ and $\theta$ directions, respectively, and $K$ is a slip coefficient; $K$ is related to a thermo-osmotic coefficient $\chi$ \cite{Bregulla2016,Fraenzl2022} as $\chi =-K T_f|_{r=a}$. 
Note that $\et\cdot \nabla T_f = \et \cdot \nabla T_s$ at $r=a$. 
Assuming that the Reynolds number $\Re$ is small, i.e., $\Re\ll 1$ (e.g., a kinematic viscosity of $\sim 10^{-6}$~m$^2$~s\mi{1} leads to $\Re \sim 10^{-5}$), the flow field $\bm{u}$ satisfies the Stokes equation, which can be solved explicitly under the boundary condition \eqref{eq:utheta_bc} and the condition imposed at infinity, namely, $\bm{u}=0$ at $r\to\infty$. The results are
\begin{subequations}\label{eq:sol_stokes}
\begin{align}
&\frac{u_r}{\alpha K} =  
 \xi_s \left(-1+\frac{a^2}{r^2}\right) \frac{a}{r}\cos\theta, \label{eq:ur}\\
&\frac{u_\theta}{\alpha K} =  
 \xi_s \left(1+\frac{a^2}{r^2}\right) \frac{a}{2r}\sin\theta, \label{eq:ut}\\
&\frac{p}{\alpha K (\eta/a) }=
- \xi_s \frac{a^2}{r^2}\cos\theta, \label{eq:p}
\end{align}
\end{subequations}
where $p$ is the pressure, $\alpha K$ is the product of the temperature gradient and the slip coefficient, namely, the characteristic speed of thermally-induced flow; $\eta$ is the viscosity of the fluid. 
In particular, on the sphere surface, we have
\begin{align}
u_\theta = - u_s \sin \theta, \quad u_s \equiv  - \xi_s K \alpha \quad (r=a),  \label{eq:u_on_boundary}
\end{align}
where the sign of $u_s$ is chosen so that $u_s$ is positive when the slip flow is induced in the negative $\theta$ direction, i.e., the hotter side of the target particle. 
When the thermal conductivity satisfies $\kappa_s=\kappa_f$, it leads to $\xi_s=1$ and thus $u_s=-K \alpha$. On the other hand, when the thermal conductivity of the target is huge, e.g., the case of a metal particle, we have $\kappa_s \gg \kappa_f$ that yields $\xi_s \ll 1$, resulting in no slip flows on the sphere surface.

\subsection{Target particle}\label{sec:target}

The fluid force $F_i$ $(i=x,\,y,\,z)$ acting on the target particle is obtained from the surface integral 
$F_i = \int p_{ij}n_j \mathrm{d}S$, where $p_{ij}$ denotes the stress tensor for a Newtonian fluid, $n_i$ is a unit vector on the surface pointing to the fluid, $\mathrm{d}S$ is an area element, and the integral is carried out over the sphere surface $r=a$. 
The $x$ component $F_x$ is then readily obtained as 
$F_x = 4\pi a\eta \xi_s K \alpha$ using Eq.~\eqref{eq:sol_stokes}. Therefore, in this model, the $x$ component of the thermophoretic velocity of the target particle, $v_{T}^\targ$, is computed from the balance between the force $F_x$ and the Stokes drag $6\pi a \eta v_{T}^\targ$, that is, 
$v_{T}^\targ=(2/3) \xi_s K \alpha$ $[=-(2/3)u_s]$. 
This means that the positive $u_s$ (i.e., a slip flow toward the hot) induces the motion of the target particle toward the cold, $v_T^\targ < 0$.

Recall that, in general, the thermophoretic velocity $\bm{v}_T$ of particles is given by $\bm{v}_T=-D_T\nabla T_f$ phenomenologically, where $D_T$ is called a thermophoretic mobility \cite{Piazza2008,Piazza2008a}. In this equation, the term $\nabla T_f$ should be understood as the temperature gradient at infinity, that is, the effect of temperature deformation due to the presence of the target particle is confined in the mobility $D_T$. Therefore, by noting that $v_T^\targ = \bm{e}_x \cdot \bm{v}_T$ and $\bm{e}_x\cdot \nabla T_f = \alpha$ with $\bm{e}_x$ a unit vector in the $x$ direction, the thermophoretic mobility of the target, $D_T^\targ$, is obtained as, 
\begin{align}
D_T^\targ \equiv -\frac{v_T^\targ}{\alpha} = -\frac{2}{3}\xi_s K, \label{eq:DTtarg}
\end{align}
which is the relation between $D_T^\targ$ and the slip coefficient $K$. 
Experimentally, the slip coefficient $K$ can be estimated from \eqref{eq:DTtarg} through the evaluation of $D_T^\targ$, i.e., $v_T^\targ$ and $\alpha$.

\subsection{Tracers}

Now, suppose that the tracers are confined in the circular path $r=R$ because of the trapping laser (see, Fig.~\ref{fig:schematic}(c)). 
The number density of the tracers is denoted by $c$. 
The velocity of the tracer, $\bm{v}$, is the sum of the flow velocity $\bm{u}$ and the thermophoretic velocity $\bm{v}_{T}^\trac\equiv -D_T^\trac \nabla T_f$ with $D_T^\trac$ the thermophoretic mobility of the tracer. 
In this equation, $\nabla T_f$ should be interpreted as a local quantity due to the smallness of the tracer, i.e., $\nabla T_f$ is computed from \eqref{eq:temperature_fluid}. 
Then, the tracer distribution $c$ satisfies a steady drift-diffusion equation with a drift velocity $\bm{v} \equiv \bm{u}+\bm{v}_T^\trac$. The $\theta$ component of $\bm{v}$ can be computed explicitly as $\bm{v}\cdot\et=- v^\ast \sin\theta$, where $v^\ast$ is the flow magnitude in the direction toward hotter region, i.e., 
\begin{align}
v^\ast 
&=  -\xi_s\left(1+\frac{a^2}{r^2}\right)\frac{a}{2r}\alpha K-\left(1+\frac{a^3}{r^3}\xi_f\right)\alpha D_T^\trac. 
\label{eq:vast}
\end{align}

Because of the scanning trapping laser at $r=R$, we assume that the tracer's radial motion is suppressed and only the motion in the $\theta$ direction is significant. 
Further assuming that there is no steady flux of the tracers,
the drift-diffusion equation is reduced to the balance between the drift and diffusion in the $\theta$ direction at $r=R$: 
\begin{align}
c (\bm{v}\cdot \et) - \frac{D}{R}\rnd{c}{\theta} = 0, 
\label{eq:drift-diffusion}
\end{align}
where $D$ is the diffusion coefficient of the trapped tracers. 
Equation~\eqref{eq:drift-diffusion} can be solved explicitly as 
\begin{subequations}\label{eq:c}
\begin{align}
&\frac{c}{c_0}= \exp(\gamma \cos\theta), \\ 
&\gamma \equiv \frac{v^\ast|_{r=R}}{D/R}, \label{eq:gamma}
\end{align}    
\end{subequations}
where $c_0$ is a normalization constant and the pre-factor $\gamma$ \eqref{eq:gamma} is the ratio between the magnitude of the drift velocity $v^\ast$ and the diffusive speed $D/R$. 
Note that $\gamma$ depends on the radius of the circular path $R$, the thermophoretic mobility $D_T^\targ$ (or $K$) and $D_T^\trac$, and the temperature gradient $\alpha$. 
When $\gamma > 0$ (or $\gamma<0$), the number density of the tracers $c$ takes maximum at $\theta=0$ (or $\pi$), i.e., the hotter (or colder) side of the target particle. When $|\gamma|\approx 0$, the diffusion dominates and $c$ tends to be distributed uniformly on the circular path. 

Model \eqref{eq:c} is used to fit and analyze the experimental results. Technical details of the fitting procedure are given in the Appendix~\ref{sec:data}. 
Once $\gamma$, $D_T^\trac$, and $\alpha$ are obtained experimentally, one can evaluate the slip coefficient $K$ using Eqs.~\eqref{eq:vast} and \eqref{eq:gamma}. 
Later, we will compare $K$ thus obtained with that estimated in the method described in Sec.~\ref{sec:target} previously. 

\section{\label{sec:res}Results and Discussion}
\subsection{Control experiments}
Before going into the main part of experimental results, we show the results of some control experiments.





\begin{figure}[tb]
\begin{center}
\iffigure
\includegraphics[width=1\linewidth]{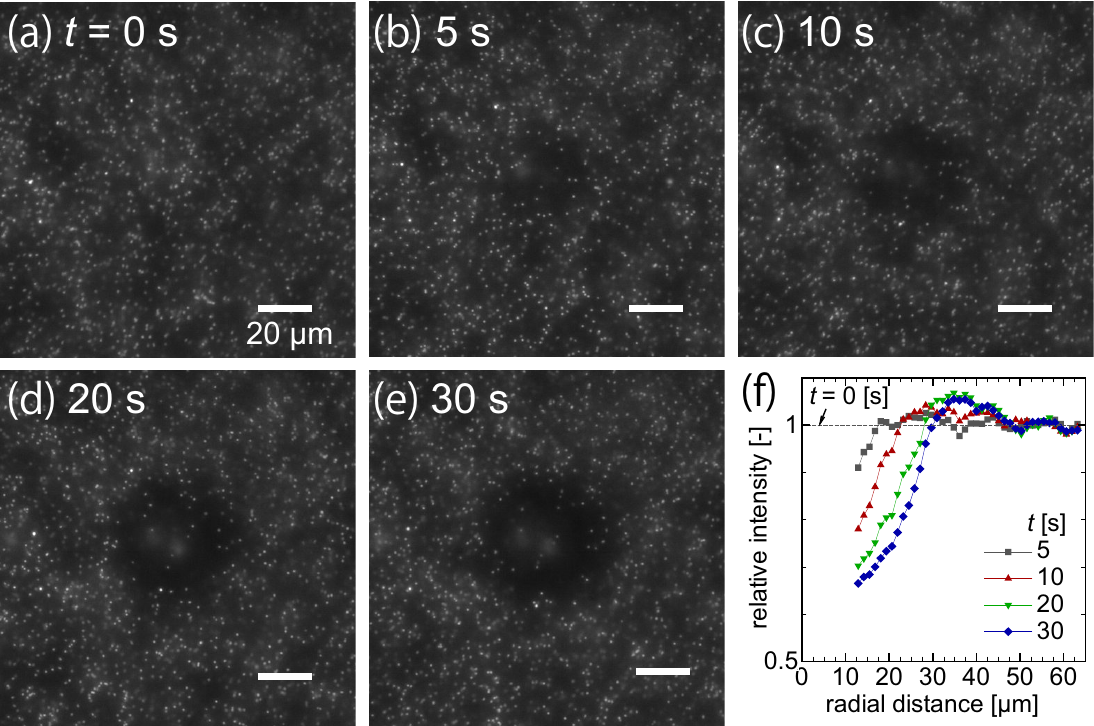}
\fi
\caption{
Thermophoresis of tracers. Snapshots at (a) $t=0$, (b) $5$, (c) $10$, (d) $20$, and (e) $30$~s, where a heating laser with $P=13.5$~mW is irradiated at the center of images. 
(f) Relative intensity distribution at $t=5$, $10$, $20$, and $30$~s averaged over $5$~s as the function of the distance $r^\prime$ from the laser. 
}\label{fig:thermo}
\end{center}
\end{figure}

\subsubsection{Thermophoresis of the tracers}\label{sec:thermo}
Firstly, we observe the thermophoretic motion of tracers in the absence of the target particle. 
In the previous studies, the tracers are known to move against the temperature gradient with the present experimental condition \cite{Tsuji2018b,Tsuji2021}. That is, $D_T^\trac$ is positive. 

Figure~\ref{fig:thermo}(a--e) shows snapshots of the thermophoresis of tracers (whites dots) near the upper surface of the microchannel at time $t=0$, $5$, ..., $30$~s, where the heating laser with a power $P=13.5$~mW is irradiated at the center of the images. 
We observe two characteristic features in the figure: (i) the fluorescent intensity increases near the radial distance $r^\prime\approx0$~\textmu m, where $r^\prime$ is the distance from the heating laser, and (ii) the dark region with low intensity spreads over $r^\prime\gtrsim 10$~\textmu m.

The former feature (i) is attributed to the tracers accumulated at the bottom surface. 
Such accumulation is due to thermophoresis in the negative $z$ direction \cite{Duhr2005}, which occurs due to a cooling at the bottom glass substrate with a rather high thermal conductivity compared with water and/or the upper surface made of polydimethylsiloxane (PDMS). These are out of the scope of the present paper. 

On the other hand, the latter feature (ii) indicates the onset of the thermophoresis of tracers in the radial direction. 
Figure~\ref{fig:thermo}(f) shows the time-development of the normalized intensity distribution in panels (a--e), and the depletion of the tracer, i.e., positive $D_T^\trac$, is clearly presented.






\begin{figure}[tb]
\begin{center}
\iffigure
\includegraphics[width=\linewidth]{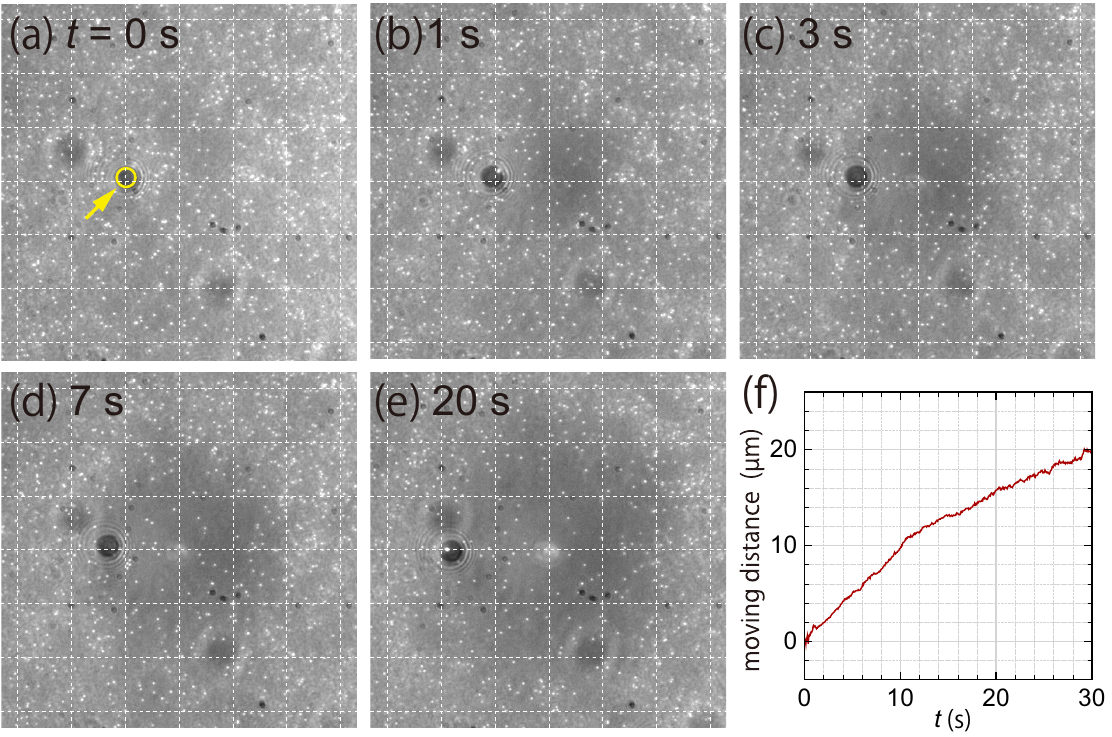}
\fi
\caption{
Bright field imaging of the thermophoresis of a target particle with a carboxylate-modified surface. Snapshot at (a) $t=0$, (b) $1$, (c) $3$, (d) $7$, and (e) $20$~s, where a heating laser is irradiated at the center of images. 
The target particle is indicated by an arrow and a yellow circle with a diameter of 7~\textmu m in panel (a) for eye guide. Grid spacing is 20~\textmu m.
(f) Moving distance as a function of time $t$. 
}\label{fig:PS_thermo}
\end{center}
\end{figure}

\subsubsection{Thermophoresis of the target particle}\label{sec:PS_thermo}
Figure~\ref{fig:PS_thermo} shows the thermophoresis of the target particle before immobilization. 
Here and in what follows, the target has carboxylate-surface modification unless otherwise stated. 
Note that a target, without immobilization, settles out onto the bottom surface of the microchannel because of its large sedimentation velocity ($>1$~\textmu m~s\mi{1}).

As shown in Fig.~\ref{fig:PS_thermo}(a) at $t=0$~s, the target particle is placed at a distance $r^\prime = 19.4$~\textmu m away from the heating laser. As time goes on (see panels (b--e)), the target particle is repelled from the heated spot, moving over almost $20$~\textmu m in the duration of $20$~s. 
Figure~\ref{fig:PS_thermo}(f) shows the moving distance of the target particle as a function of time, obtained using the particle tracking analysis of an open-source image-processing software, ImageJ. 

The average speed of the target particle up to $t=2$~s is $0.89$~\textmu m~s$^{-1}$ and the average temperature and temperature gradient over the trajectory is $328$~K and $-1.77\pm0.08$~K~\textmu m\mi{1}, respectively, leading to the thermophoretic mobility of the target, $D_T^\targ = 0.50\pm 0.17$~\textmu m$^2$~s$^{-1}$~K$^{-1}$.





\subsubsection{Thermophoresis of the trapped tracers}
\begin{figure}[tb]
\begin{center}
\iffigure
\includegraphics[width=\linewidth]{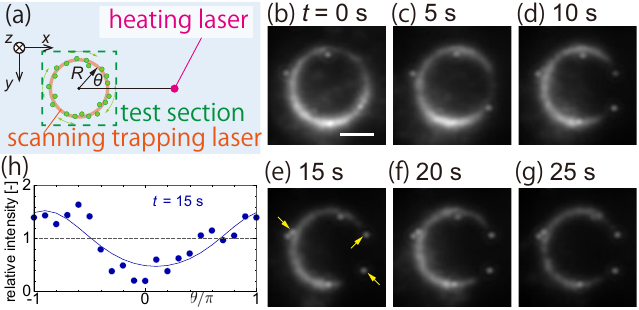}
\fi
\caption{
(a) Schematic of an experiment in the absence of a target particle. (b--g) Time-averaged images over 5~s. Arrows in panel (e) indicates tracers stuck to the channel wall. These stuck tracers can be noise in the intensity analysis. 
Scale bar: $4$~\textmu m. 
(h) Relative intensity profile at $r=R$ with respect to $\theta$.
}\label{fig:NoPS}
\end{center}
\end{figure}

In this section, we show the tracer motion which are trapped on the circular path in the absence of the target particle, as shown in Fig.~\ref{fig:NoPS}(a).
This is a control experiment to assure that the irradiation of the scanning trapping laser does not affect the direction of thermophoresis of tracers. 
The distance between the center of the scanning trapping laser and the heating laser is set to $L=15.5$~\textmu m and the laser power of the trapping laser is set to $71$~mW.

Figure~\ref{fig:NoPS}(b--g) are time-averaged images of the tracer with the radius of the circular path $R=4.4$~\textmu m. 
For instance, panel (b) is the time-averaged images over $-5$~s $\leq t\leq 0$~s. Since these are time-averaged images, if a tracer looks not blurry, it is not a Brownian particle, that is, the tracer adheres to a channel wall (see, e.g., the tracers indicated by arrows in panel (e)). 

It is seen that the tracers near $\theta=0$ are depleted as time goes on, i.e., the intensity reduces near $\theta=0$. Moreover, the overall intensity decreases, indicating that some tracers escape from the optical trapping and move away from the hot region. Figure~\ref{fig:NoPS}(h) shows the relative intensity distribution near $r=R$ and $t=15$~s (i.e., panel (e)) as a function of $\theta$. The fitting is made by Eq.~\eqref{eq:c} with $\gamma=-0.52$ (see Appendix~\ref{sec:data}), indicating the motion of tracers toward the colder side. 
In the absence of the target particle, the temperature gradient is uniform (see, Eq.~\eqref{eq:temperature_fluid}) and the slip flow is absent. Therefore, $|\gamma|$ corresponds to the phoretic P\'eclet number \cite{Mayer2023}, that is, the magnitude $|\gamma|\approx O(1)$ indicates that the thermophoresis of the tracers is competing with the diffusion of themselves. 
Noting that $\xi_f = 0$ (or $\kappa_f = \kappa_s$) and $K=0$ without the target particle, the value of $\gamma=-0.52$ results in $D_T^\trac=0.10$~\textmu m$^2$~s\mi{1}~K\mi{1} using Eqs.~\eqref{eq:gamma} and \eqref{eq:vast}.


\begin{figure*}[tb]
\begin{center}
\iffigure
\includegraphics[width=0.8\linewidth]{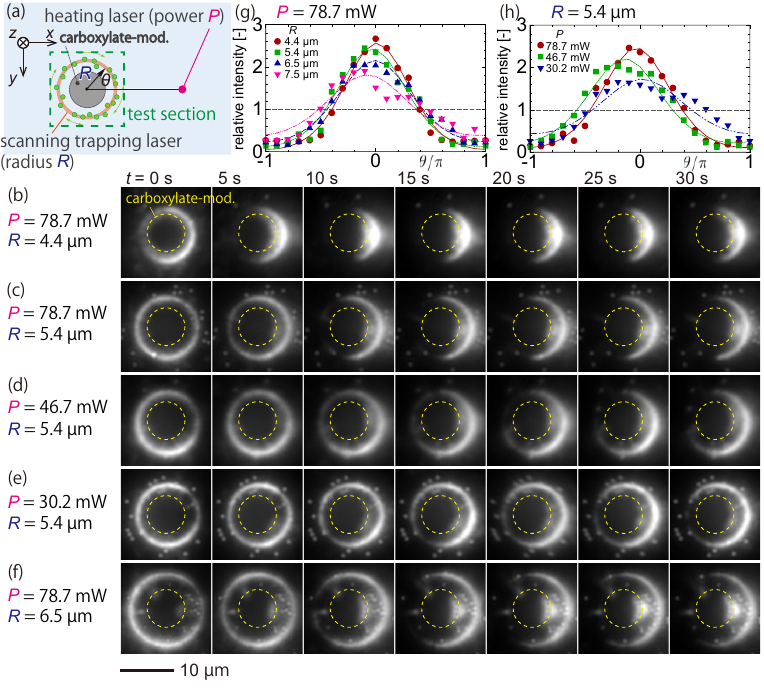}
\fi
\caption{(a) Schematic of experiments in the presence of a target particle with a carboxylate-modified surface ($\zeta=-46.0$~mV). 
$R$ is the radius of a scanning trapping laser and $P$ is the power of a heating laser. (b--f) Time-averaged images over 5~s for various $P$ and $R $: (b,c,f) $P=78.7$~mW, (d) $P=46.7$~mW, and (e) $P=30.2$~mW; 
(b) $R=4.4$~\textmu m, (c--e) $R=5.4$~\textmu m, and (f) $R=6.5$~\textmu m. Since these are time-averaged images, if a tracer looks not blurry, it adheres to the upper channel wall. 
(g,h) Relative intensity profile as a function of $\theta$; (g) $P=78.7$~mW ($t=25$~s) for various $R$ and (h) $R=5.4$~\textmu m for various $P$ ($t=30$~s). 
See also a movie``Movie-Fig5(b)-carboxylate-mod.avi" for panel (b).}
\label{fig:main}
\end{center}
\end{figure*}

\subsection{Observation of thermally-induced flows} \label{sec:main}
Let us describe the main result of the present paper, the thermally-induced flow around the target particle. 
Firstly, the results for a target particle with a carboxylate-modified surface are presented. 
The zeta potential $\zeta$ of target particles is measured from the electrophoretic mobility (ELSZ-2000Z, Otsuka Electronics) as $\zeta=-46$~mV.
Figure~\ref{fig:main}(a) shows the schematic of the situation presented in panels (b--f), which are the recorded images for various parameters $P=78.7$. $46.7$, and $30.2$~mW (the power of the heating laser) and $R=4.4$, $5.4$, $6.5$, and $7.5$~\textmu m (the radius of the scanning trapping laser). One can compare the effect of $R$ by inspecting panels (b,c,f) and that of $P$ by panels (c,d,e). 
Lowering the laser power to $P=13.5$~mW did not show a significant change in the tracer distribution from the initial state.

Firstly, it is seen from panel (b) that the tracers, which are initially rather uniform in the $\theta$ variable, tend to gather toward the hotter side (i.e., the right side, see panel (a)) as time goes on. This tendency is the same for other panels (c--f), and should be contrasted with Fig.~\ref{fig:NoPS} for a case without a target particle; the tracers gather toward the colder side. 

As can be seen from the comparison between panels (b) $R=4.4$~\textmu m, (c) $R=5.4$~\textmu m, and (f) $R=6.5$~\textmu m, the tracers' movement is more intense and faster for smaller $R$. 
To be more precise, at time $t=5$~s, panel (b) starts to exhibit the depletion of the tracers near $\theta=\pi$ (the colder side; see panel (a)); however panels (c) and (f) show the apparent depletion near $\theta=\pi$ only for $t\geq 10$~s. 
At $t=30$~s, where the distributions can be considered steady, the intensity near $\theta=0$ seems to reach maximum, indicating the accumulation near the hot side. 
To compare more quantitatively, we show in panel (g) the relative intensity distribution (symbols) and the fitting using Eq.~\eqref{eq:c} (lines) (see also Appendix~\ref{sec:data}). 
Although the results are qualitative, the plots in panel (g) capture the observation made in panels (b,c,f). 

It should be noted that some tracers seem trapped in the region between the upper wall of the microchannel and the target particle (see the region ``$\mathsf{A}$" in Fig.~\ref{fig:schematic}(b)), as clearly seen in Fig.~\ref{fig:main}(f) at $t\geq 10$~s. 
These tracers can stay there even when the scanning trapping laser is not irradiated. We suppose that the thermo-osmotic flow and the thermophoresis in the $z$ direction may cause this trapping, but further discussion is difficult using only the present two-dimensional information on the $x\,y$ plane, and thus future work. 

One can compare the effect of the heating laser power $P$ by seeing 
(c) $P=78.7$~mW, (d) $P=46.7$~mW, and (e) $P=30.2$~mW. 
The corresponding temperature-gradient measurement given in Appendix~\ref{sec:temp} shows that $|\nabla T_f|$ is approximately 
(b) $2.1$~K~\textmu m\mi{1}, 
(c) $1.6$~K~\textmu m\mi{1}, and
(d) $1.0$~K~\textmu m\mi{1} near $r^\prime \approx L=18$~\textmu m, at which the target particle is fixed. 
Therefore, the tracer movement is more intense for the case with larger $P$. 
Figure~\ref{fig:main}(h) shows the corresponding relative intensity distribution at $t=30$~s. 
Sharper shape near $\theta=0$ for larger $P$ indicates the more intense accumulation of tracers near the hotter side.

Next, let us describe the thermally-induced flow around a target particle with an amine-modified surface, the zeta potential $\zeta$ of which is measured as $\zeta=-23.7$~mV. 
The thermophoretic mobility of a target particle with the amine-modified surface is estimated in the same manner as Sec.~\ref{sec:PS_thermo}, and we obtain $D_T^\targ=0.29\pm0.08$~\textmu m$^2$~s\mi{1}~K\mi{1} where the average temperature and temperature gradient are $320$~K and $-1.18$~K~\textmu m\mi{1}, respectively.
This value of $D_T^\targ$ is smaller than that of a target particle with the carboxylate-modified surface (see Sec.~\ref{sec:PS_thermo}). 

\begin{figure}[tb]
\begin{center}
\iffigure
\includegraphics[width=\linewidth]{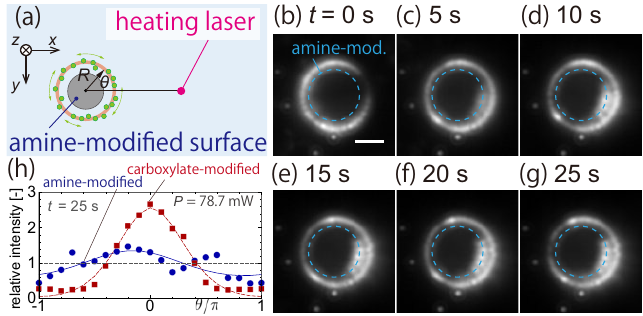}
\fi
\caption{
(a) Schematic of experiments in the presence of a target particle with an amine-modified surface ($\zeta=-23.7$~mV).  (b--g) Time-averaged images over 5~s. 
Scale bar: $4$~\textmu m. 
(h) Relative intensity profile at $r=R$ with respect to $\theta$ for amine-modified (circle) and carboxylate-modified (square; see also Fig.~\ref{fig:main}(b) for microscope images). 
See also a movie``Movie-Fig6(b-g)-amine-mod.avi" for panel (b--g).}
\label{fig:Main-NH2}
\end{center}
\end{figure}

Figure~\ref{fig:Main-NH2}(a) shows the schematic of the situation presented in panels (b--g), which are recorded images for $P=78.7$~mW and $R=4.4$~\textmu m with a target particle with the amine-modified surface. See Fig.~\ref{fig:main}(b) for comparison with the case of a carboxylate-modified surface. 
Remarkably, the case of the amine-modified target results in a faint accumulation of tracers. As shown in panels (b-g), the tracers slightly gather toward the hotter side, but the accumulation is very weak compared with Fig.~\ref{fig:main}(b) for a carboxylate-modified target. 
This observation is further verified by the profile of the relative intensity distribution shown in panel (h) for both amine (circle) and carboxylate (square) modified surfaces. 
Since the surface modification is the only difference between these two cases, panel (h) is a clear evidence that the observed tracer movement originates from surface phenomena. 


\subsection{Discussion}

To compare the experimental results with existing models of thermally-induced slip flows, we use the model proposed in Ref.~\cite{Fayolle2008}, in which a thin electric-double-layer (EDL) approximation for a charged surface of a spherical colloid is applied under the Debye-H\"uckel approximation. The resulting slip velocity is written as
\begin{subequations}\label{eq:vB}
\begin{align}
&v_B = - v_0 \sin\theta \quad (r=a), \\
&v_0 \equiv \frac{\varepsilon \psi_0^2}{8 \eta} \left(1 - \frac{\mathrm{d} \ln \varepsilon}{\mathrm{d}\ln T_f}\right)  \xi_s 
\frac{|\nabla T_f|_{\infty}}{T_f},  
\end{align}
\end{subequations}
where $|\nabla T_f|_{\infty}|(=\alpha)$ is the temperature gradient at infinity,  $\varepsilon$ is the temperature-dependent electric permittivity, and $\psi_0$ is the electrostatic potential at the surface. 
We assume that $\psi_0\approx \zeta$. 
The temperature dependence of $\varepsilon$ for water is modeled as $\varepsilon/\varepsilon_0=305.7\times \exp(-T_f/217)$ \cite{Xuan2004} 
where $\varepsilon_0=8.85\times10^{-12}$~F~m\mi{1} is the electric permittivity in vacuum. 
Furthermore, the empirical model of viscosity $\eta= (2.761 \times 10^{-6}) \times \exp(1713/T_f)$~Pa~s \cite{Knox1994} is used. 
For the above model, the slip coefficient $K$ in Eq.~\eqref{eq:u_on_boundary} is expressed as $K=-v_0/(\xi_s \alpha)$ because $u_s=v_0$. To be more specific, we have
\begin{align}
K = - \frac{\varepsilon \zeta^2}{8\eta T_f} \left(1 - \frac{\mathrm{d} \ln \varepsilon}{\mathrm{d}\ln T_f}\right). \label{eq:K-PRE}
\end{align}

In Ref.~\cite{Bregulla2016}, a similar form is proposed based on the Derjaguin’s model \cite{Derjaguin1965} with an enthalpy excess given by EDL enthalpy with the Debye-H\"uckel approximation. 
This model was considered to analyze the thermo-osmosis on a bare glass substrate. The result was 
\begin{align}
K=-\frac{\chi}{T_f} = -\frac{\varepsilon \zeta^2 }{8 \eta T_f}. 
\label{eq:K-PRL}
\end{align}

The theoretical prediction $K$ obtained by Eqs.~\eqref{eq:K-PRE} and \eqref{eq:K-PRL} can be compared with $K$ obtained from the experiments via two approaches.   
(Method 1) First, $K$ can be obtained from the thermophoretic motion of the target, Eq.~\eqref{eq:DTtarg}, because we have obtained $D_T^\targ$  experimentally by the particle tracking. See Sec.~\ref{sec:PS_thermo}. 
(Method 2) Second, $K$ can be obtained from the observation of slip flows in Sec.~\ref{sec:main}, namely, the tracer accumulation behavior \eqref{eq:c}, through the fitting of the value $\gamma$ (see Figs.~\ref{fig:main} and \ref{fig:Main-NH2} for the result of fitting). 
Then, equation~\eqref{eq:vast} and $\gamma$ yield $K$ since we have already evaluated $D_T^\trac$ in Sec.~\ref{sec:thermo}.

\begin{figure}[tb]
\begin{center}
\iffigure
\includegraphics[width=\linewidth]{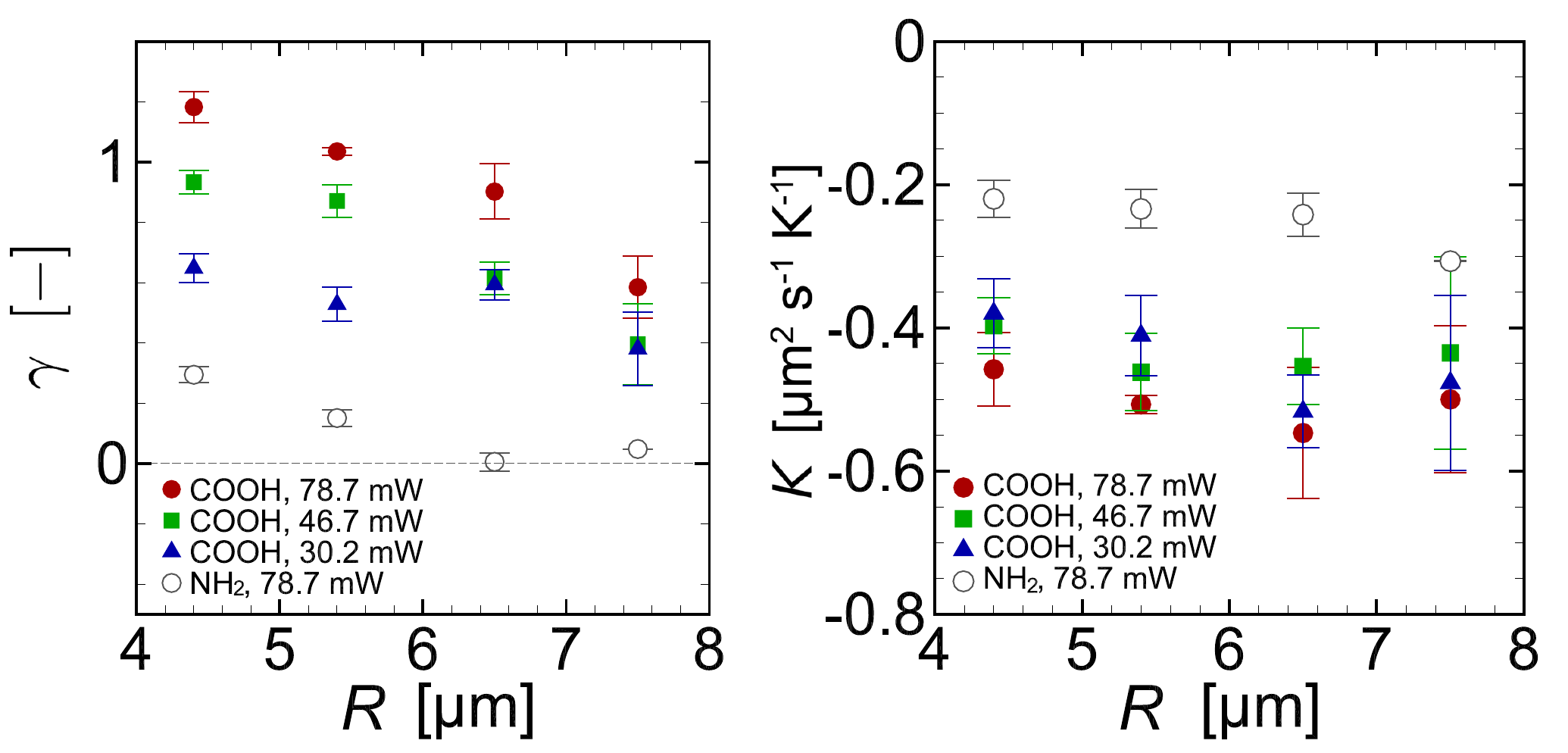}
\fi
\caption{(a) The values of $\gamma$ as a function of a scanning radius $R$ for carboxylate-modified (COOH, filled symbols) and amine-modified (NH$_2$, empty symbols) surfaces with various heating power $P=78.7$ (circle), $46.7$ (square), and $30.2$~mW (triangle). (b) Slip coefficient $K$ computed from $\gamma$ in panel (a) through Eqs.~\eqref{eq:gamma} and $\eqref{eq:vast}$. 
}\label{fig:PREcompare}
\end{center}
\end{figure}

Before going into the comparison between Eq.~\eqref{eq:K-PRE}, Eq.~\eqref{eq:K-PRL}, Method 1, and Method 2, we show the values of $\gamma$ used in Method 2. 
Figure~\ref{fig:PREcompare}(a) is the values of $\gamma$ averaged from $t=10$~s to $30$~s for the heating laser power $P=78.7$~mW (circle), $46.7$~mW (square), and $30.2$~mW (triangle) as a function of the scanning radius $R$. Filled symbols indicate the results of the carboxylate-modified target particles and empty symbols those of the amine-modified target particles. 
Although qualitative, we see that $\gamma$ tends to decrease as $R$, that is, the magnitude of thermally-induced flow decreases. 
Moreover, comparison over different laser power $P$ shows that $\gamma$ decreases with $P$. 
The cases of amine-modified target yield significantly smaller $\gamma$ than those of the carboxylate-modified target. 
These values of $\gamma$ can be converted into $K$ through Eqs.~\eqref{eq:gamma} and \eqref{eq:vast}; the results are summarized in Fig.~\ref{fig:PREcompare}(b). 
For the case of carboxylate-modified surface, $K$ seems to take the value around $-0.4$~\textmu m$^2$~s\mi{1}~K\mi{1}, while that of amine-modified surface results in
$K\approx -0.2$~\textmu m$^2$~s\mi{1}~K\mi{1}. 

Finally, we summarize in Table~\ref{tab:K} the results of $K$ obtained by Eq.~\eqref{eq:K-PRE}, Eq.~\eqref{eq:K-PRL}, Method 1, and Method 2, where the average over different $R$ is taken for Method 2 (see Fig.~\ref{fig:PREcompare}(b)). 
Furthermore, Table~\ref{tab:K-cichos} shows the values of $K$ computed from the existing experimental results of thermo-osmotic flows \cite{Bregulla2016,Fraenzl2022} with different surface details. 
All the values of $K$ in both tables are negative, i.e., the flow is induced from cold to hot side. The magnitude of $K$ in Table~\ref{tab:K} has the same order $O(1)$~\textmu m$^2$~s\mi{1}~K\mi{1} but quantitative agreement is not obtained between theory and experiments. 
However, importantly, Methods 1 and 2 show similar values of $K$ and capture the difference of surface modification predicted by Eqs.~\eqref{eq:K-PRE} and \eqref{eq:K-PRL}: $K$ for the carboxylate-modified case is larger than that for the amine-modified case.
Comparing Tables~\ref{tab:K} and \ref{tab:K-cichos}, it is seen that the slip coefficients obtained in the present experiments are close to that for bare glass. 

The thermo-osmotic coefficients have been also computed using molecular dynamics simulation in Refs. \cite{Fu2017} and \cite{Chen2021a} for Lennard-Jones fluid/solid system and water/quartz system, respectively. 
The results were $\chi\sim O(10^{-8})$~m$^2$~s\mi{1} and one or two orders magnitude larger than theory and experiments. 

The reason for the quantitative discrepancy between theory and experiment cannot be fully clarified at the moment, but let us point out some possibilities. 
First, Eq.~\eqref{eq:K-PRE} was derived under the thin-EDL and Debye–H\"uckel approximation. The former is valid because we use an electrolyte solution with 10~mM ionic concentration. But the latter may be violated due to the non-negligible magnitude of the zeta potential compared with $k_B T_f$. 
Experiments, i.e., Method 1 or Method 2, may underestimate $v_T^\targ$ or $K$ because the target particle stays close to the microchannel walls due to a technical limitation in the experiment. Further improvement of the proposed method and its application to other systems is future work. 

\begin{table*}[tb]
    \centering
    \caption{Slip coefficient $K$ [\textmu m$^2$~s\mi{1}~K\mi{1}] obtained from the model \eqref{eq:K-PRE} and \eqref{eq:K-PRL} based on Refs.~\cite{Fayolle2008,Bregulla2016} and the present experimental results with Method 1 and Method 2.}
        \begin{tabular}{lccccccccc}
     \hline
                                          & Eq.~\eqref{eq:K-PRE}    &~~& Eq.~\eqref{eq:K-PRL}     &~~& Method 1     &~~& \multicolumn{3}{c}{Method 2} \\
                                          & Ref.~\cite{Fayolle2008} &  & Ref.~\cite{Bregulla2016} &  &              &~~& $P=78.7$~mW    & $P=46.7$~mW    & $P=30.2$~mW   \\
                                          \cline{2-2} \cline{4-4} \cline{6-6}  \cline {8-10}
     carboxylate-mod. ($\zeta=-46.0$~mV) & $-2.36$                 &  & $-0.94$                  &  &$-0.58\pm0.20$&~~& $-0.50\pm0.02$ & $-0.44\pm0.02$ & $-0.45\pm0.03$\\
     amine-mod. ($\zeta=-23.7$~mV)       & $-0.58$                 &  & $-0.23$                  &  &$-0.34\pm0.10$&~~& $-0.25\pm0.01$ &                &               \\
     \hline
    \end{tabular}
    \label{tab:K}
\end{table*}

\begin{table*}[tb]
    \centering
    \caption{Slip coefficient $K=-\chi/T_f$ [\textmu m$^2$~s\mi{1}~K\mi{1}] reported in Ref.~\cite{Bregulla2016} for a glass substrate with or without non-ionic copolymer surfactant (Pluronic F-127) coating and Ref.~\cite{Fraenzl2022} for a thin gold film on a glass. In the references, the thermo-osmotic coefficient $\chi$ is reported and the values shown in this table are obtained for $T_f=300$~K and $330$~K.}
    \begin{tabular}{lccc}
     \hline
      Wall condition (experiments) & $\chi$~[m$^2$~s\mi{1}]& $K$ at $T_f = 300$~K & $K$ at $T_f = 330$~K\\
      \hline
      Glass coated with Pluronic F-127 \cite{Bregulla2016}  & $13 \times 10^{-10}$ & $-4.3$&$-3.9$\\
      Bare glass \cite{Bregulla2016}                       & $1.8\times 10^{-10}$ & $-0.6$&$-0.5$\\ 
      Gold-thin film on a glass \cite{Fraenzl2022}       & $10\times 10^{-10}$  & $-3.3$&$-3.0$\\
     \hline
    \end{tabular}
    \label{tab:K-cichos}
\end{table*}

\section{\label{sec:conc}Conclusion}
In this paper, we proposed an experimental method to characterize thermo-osmotic slip flows around a microparticle. 
The method utilized the optical trapping of tracers around a microparticle, which indicated the generation of thermally-induced flows near the particle surface through the change of tracer's spatial distribution. 
We applied the proposed method to the flow characterization around a microparticle under a strong temperature gradient created by the optical heating of a water solution.

The slip coefficient was estimated experimentally by using two methods: a thermophoretic velocity measurement of a target particle (Method 1) and a thermo-osmotic flow measurement around a target particle (Method 2), where the two methods were related through a simple physical model assuming a thermally-induced slip boundary condition. 
The values of slip coefficients obtained by the two methods were in reasonable agreement, that is, 
we demonstrated that a flow was indeed induced along the surface of the particle under the temperature gradient and the agreement provides an evidence that the flow was the major driving mechanism of thermophoresis. 
Furthermore, the experiments captured the difference of the surface modification predicted by the theory of charged surface \cite{Fayolle2008} with temperature variation.

The use of the optical trapping of tracers, as demonstrated in the present paper, can also be applied to other flows near surfaces, such as those along the surface of a heated Janus particle. Investigating not only the motion of active matter but also the induced flows around them will explore a further understanding of nanofluidics and their novel applications.  

\begin{acknowledgments}
This work was partly supported by JSPS KAKENHI grant No.~JP18H05242, JP20H02067, and JP22K18770, JP22K03924, and also by JST PRESTO grant No.~JPMJPR22O7.
\end{acknowledgments}

\appendix
\begin{figure*}[tb]
\begin{center}
\iffigure
\includegraphics[width=0.7\textwidth]{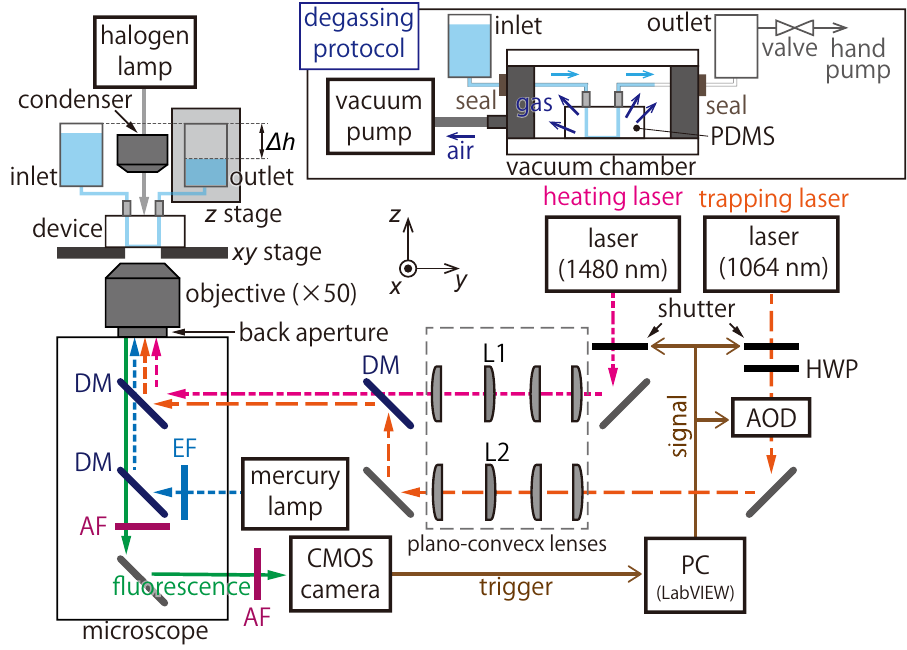}
\fi
\caption{
An overview of the experimental setup. 
AF: absorption filters, AOD: acousto-optic deflector, DM: dichroic mirrors, EF: excitation filter, HWP: half-wave plate. 
An inset shows a degassing protocol. 
}\label{fig:setup}
\end{center}
\end{figure*}
\section{Experimental details}\label{sec:exp_detail}

\subsection{Overview}
A schematic of the experiment is shown in Fig.~\ref{fig:setup}. 
A microfluidics device made of PDMS block and a cover glass is mounted on an inverted microscope (IX--73, Olympus). The fabrication process of the microchannel was given in \cite{Tsuji2018b}. The straight microchannel has a rectangular cross section with a height of $H=17$~\textmu m in the $z$ direction, a width of $400$~\textmu m in the $x$ direction and the length $>3$~mm in the $y$ direction (see also Fig.~\ref{fig:schematic}). 
The height $H=17$~\textmu m is small enough to suppress the effect of thermal convection. 

Two lasers with the wavelengths of 1064~nm (Powerwave--1064, NPI lasers) and 1480~nm (FOL1439R50--667--1480, Furukawa Electric) are irradiated to the microchannel through an objective ($\times$50, numerical aperture (NA) = 0.65; LCPLN50xIR, Olympus); 
the former/latter is called a trapping/heating laser in this paper. 
The trapping laser is scanned in the focal plane by using a two-axis acousto-optic deflector (AOD; DTD-274HD6M \& DE-272M, Intra Action) and captures tracers for flow visualization. 
The heating laser creates an inhomogeneous temperature field by using a photothermal effect of a water solution. Note that the absorption coefficient $\beta$ of the water at the wavelength of 1480~nm is $\beta=2.4\times10^3$~m$^{-1}$ \cite{Cordero2009,Riviere2016}. 
This value of $\beta$ is large enough for the necessary temperature elevation with a moderate laser power $<100$~mW. 

A background fluid flow in the microchannel in the $y$ direction is controlled by adjusting a pressure difference $\Delta P$ between inlet and outlet reservoirs (Fig.~\ref{fig:setup}). 
To be more specific, $\Delta P$ is controlled with a resolution $0.01$~Pa by the water-level difference $\Delta h$ between the two reservoirs using a $z$-stage. 
Before every experiment, we confirm that the background fluid flow is absent. 
After an experiment, there arises a non-uniform tracer distribution. 
Therefore, we induce a background flow to supply a fresh solution with a uniform tracer distribution to the test section. 

Any residue bubbles in the microchannel, connecting tubes, and/or reservoirs can lead to the loss of reproducibility of the experiments. 
To be more precise, if there are any residue bubbles, (a) the background flow never stops and there is always a very slow flow ($<1$~\textmu m) that can hinder the observation of thermally-induced flows; (b) undesired and uncontrollable tracer motion occurs just after an onset of the heating laser irradiation.
Therefore, we pay special attention to the removal of the bubbles. 
More specifically, before mounting the microfluidic device on the microscope, the whole device is degassed by putting only the PDMS-glass part in a vacuum chamber, as shown in the inset of Fig.~\ref{fig:setup}. 
Since the PDMS can absorb a gas \cite{Hosokawa2004}, the degassing can avoid the creation of bubbles and also can remove the residue bubble in the microchannel. 
By further degassing the empty outlet (see the inset of Fig.~\ref{fig:setup}) using a hand pump, we can successfully fill the whole channel (including tubes) with a sample solution without residue bubbles. 

\subsection{Sample solution}\label{sec:sample}
Target particles are non-fluorescent polystyrene spheres with a diameter of $7$~\textmu m and with carboxylate (01-02-703, Micromod) or amine (01-01-703, Micromod) surface modification. 
Tracer particles are fluorescent polystyrene spheres with yellow-green fluorescence (F8813, Molecular Probes) and have a diameter of $500$~nm and a carboxylate surface modification.
These as-purchased particles are dispersed in a tris(hydroxymethyl)amino-methane hydrochloride (tris-HCl) buffer solution ($\mathrm{pH} = 8.0$, Nippon Gene) with a concentration of 10~mM.
The concentration of the target particles and tracers are $2.4\times 10^{-4}$ \% and $4\times 10^{-2}$~\%, respectively. 
The sample solution is sonicated for 5 minutes before use to avoid the agglomeration of tracers. 

\subsection{Optical setup}
Two collimated lasers (for trapping and heating) are expanded using a set of plano-convex lenses, as shown in Fig.~\ref{fig:setup}. 
A half-wave plate (HWP) is used for the optical path of the trapping laser to adjust the direction of linear polarization to that of the specification of AOD. The lenses L1, L2, and AOD are placed at the conjugate planes of the back aperture of the objective lens, that is, the motion of L1 and L2 perpendicular to the lenses and the beam deflection at AOD can rotate the beam at the back aperture. In this way, we can move the laser focus in the focal plane in the $x$ and $y$ directions with minimal beam clipping \cite{Neuman2004}. 

Mercury lamp (U-HGLGPS, Olympus) with a band path filter (460~nm--495~nm) is used for an excitation light source in fluorescent imaging. 
A fluorescence from the tracers is collected by the same objective and is recorded by a scientific complementary-metal-oxide-semiconductor (sCMOS) camera (Zyla 5.5, Andor Technology), after removing undesired stray lights using absorption filters. 
The field depth in the $z$ direction for this observation setup is $6$~\textmu m, as measured in our previous experiments with the same objective \cite{Tsuji2021}. 
To observe the non-fluorescent target particle, we use a transmission light from a halogen lamp.



\subsection{Experimental procedure}
\begin{figure}[tb]
\begin{center}
\iffigure
\includegraphics[width=0.8\linewidth]{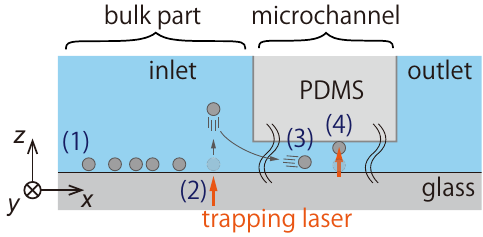}
\fi
\caption{Schematic of an immobilization process. (1) Many target particles are sedimented at the bottom of the bulk part, but (2) they are optically pushed by the laser with a wavelength of 1064 nm so that they disperse again. 
(3) The dispersed target particles are then driven into the microchannel. (4) The target particles are optically pushed to the ceil of the microchannel until they stick there. }\label{fig:immo}
\end{center}
\end{figure}

To immobilize the target particle to the upper surface in the microchannel, we use the trapping laser, following procedures (1)--(4) shown in Fig.~\ref{fig:immo}:  
(1) Target particles have a non-negligible sedimentation speed. Thus, they tend to stay at a bulk part and rarely come to the microchannel part. 
(2) We lift a target particle in the bulk by irradiating the trapping laser. 
(3) We induce a background flow to bring the target particle to the center of the microchannel part. 
(4) We again lift the target particle by irradiating the trapping laser and push it in the positive $z$ direction to the upper surface of the microchannel, waiting for about 10 minutes until the target particle sticks to the upper surface. 

Note that the $y$ position of the immobilized particle should be about the center of the microchannel. 
Otherwise, we observe very slight but non-negligible tracer motion in the $y$ direction just after the initiation of laser heating. 
We consider that this unwanted tracer motion does not originate from the residue bubbles' existence, because the magnitude is much weaker than that of the case with the residue bubbles.
In Fig.~\ref{fig:schematic}(b), we take the $y$ axis, i.e., the direction of the background flow, perpendicular to the line of $\theta = 0$ to assure that the accumulation (or depletion) of tracers (see Sec.~\ref{sec:res}) is not caused by such bias if any.

After the immobilization of the target particle, we scan the trapping laser in a circular path with the radius $R$ with a power of $178$~mW (after the objective). 
The scanning frequency is set to 200~Hz, which is high enough to avoid forcing tracers to orbit around the target particle. 
After the realization of the homogeneous distribution of tracers with respect to $\theta$, the heating laser with a power of $P$ (after the objective) is irradiated, where $P=13.5$, $30.2$, $46.7$, and $78.7$~mW are used in this paper. 
The motion of tracers are recorded with a frame rate of $20$~fps for $>25$~s.

\begin{table}[tb]
    \centering
    \caption{Parameters used in the fitting of Fig.~\ref{fig:temperature}.}
    \label{tab:temperature}
    \begin{tabular}{ccccc}
    \hline
       $P$~[mW] &\qquad & $\Delta T$~[K]&$\sigma$~[\textmu m]& $T_0$~[K]\\
       \hline
       $78.7$&&75.7&16.9&24.6\\
       $46.7$&&59.5&14.0&24.6\\
       $30.2$&&41.6&13.1&24.1\\
       $13.5$&&21.4&11.7&21.6\\
    \hline
    \end{tabular}
\end{table}

\section{Temperature profile}\label{sec:temp}
We visualize the temperature profile of the sample solution in the $x\,y$ plane using a laser-induced fluorescence method \cite{Tsuji2018a,Braun2002,Cordero2009,Riviere2016,Tsuji2021}. 
The measurement process is exactly same as that in Refs.~\cite{Tsuji2018a,Tsuji2021} and the details are omitted here. 
Note that the measured temperature profile below is the temperature of the fluid integrated over the depth (i.e., the $z$) direction.

\begin{figure}[bt]
\begin{center}
\iffigure
\includegraphics[width=\linewidth]{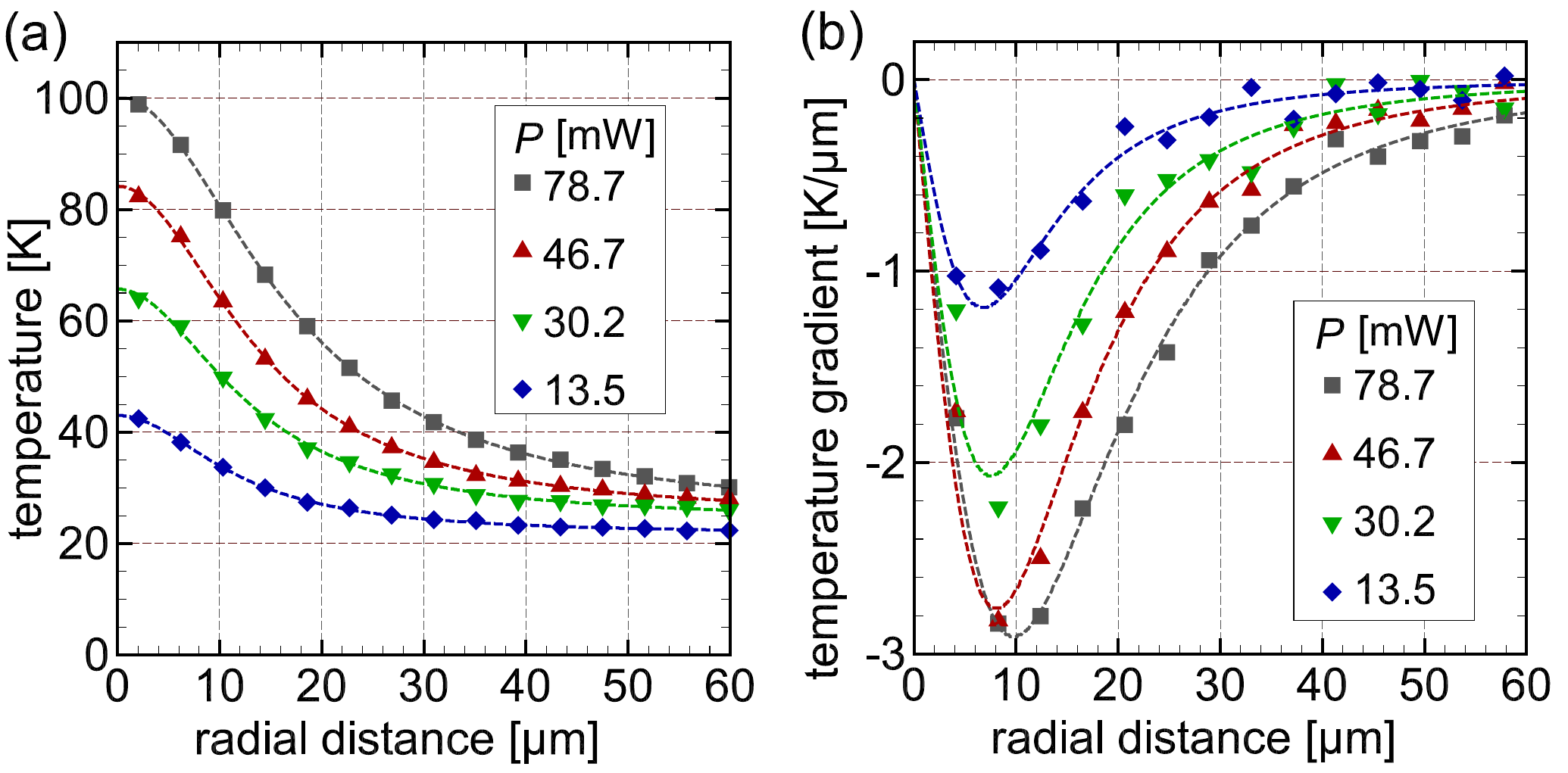}
\fi
\caption{(a) Temperature of a fluid and (b) a radial temperature gradient as a function of a radial distance $r^\prime$ from a heating spot (see, Fig.~\ref{fig:schematic}(c)). 
The power $P$ of the laser with a wavelength 1480~nm are set to $P=78.7$, $46.7$, $30.2$, and $13.5$~mW.}\label{fig:temperature}
\end{center}
\end{figure}

Figure~\ref{fig:temperature}(a) shows the result of temperature measurement for the laser power $P$ used in this paper. The experimental data are shown by symbols and we fit these plots using half Lorentzian curves, $T_0 + \Delta T/(1 + (r^\prime/\sigma)^2 )$, where $\Delta T$ is the temperature increase at the beam center, $T_0$ is the temperature at infinity, $\sigma$ is the half-width at the midheight value of the temperature profile, and $r^\prime$ is the distance from the heating spot. The parameters used in the fitting are summarized in Table~\ref{tab:temperature}. The corresponding radial gradient is also shown in Fig.~\ref{fig:temperature}(b).

\section{Data analysis procedure}\label{sec:data}
Let $f_{ij}^n$ be the intensity of an obtained image, where the indices $i$ and $j$ are the pixel numbers in the $x$ and $y$ directions; $n$ is the frame number. 
Firstly, we average the images over $n_0$ frames with $n_0=100$ (over 5~s) to reduce noise and define them as $\bar{f}^{n}_{ij}=\sum_{m=n-n_0}^n \bar{f}^{m}_{ij}$. 
Then, we subtract the background value, which is the minimum value over whole pixel of $\bar{f}^{n}_{ij}$. 
The $\theta$ direction is divided into intervals and we introduce $\theta_p = p \Delta \theta$ $(p=-p_0,...,p_0)$ with $p_0=10$ and $\Delta \theta = \pi/p_0$. 
To analyze the result with $r=R$, we average the intensity $\bar{f}^{n}_{ij}$ over a domain $(r,\theta)\in [R-\Delta r,R + \Delta r]\times[\theta_p,\theta_p+\Delta \theta]$ with $\Delta r \approx 258$~nm; 
This averaged value is named as $F_p^n$ and we then compute the spatial average over whole $\theta$, i.e., $\bar{F}^n=\frac{1}{2\pi}\sum_{\text{all }p}F_p^n \Delta \theta$. The results shown in the main text (e.g., Sec.~\ref{sec:main}) are the relative intensity, namely, $F_p^n/\bar{F}^n$ [$=$ (intensity near $\theta = \theta_p$) / (intensity averaged over whole $\theta$)].

Next, we describe details of the fitting procedure. 
In practice, the value of the normalization constant $c_0$ in Eq.~\eqref{eq:c} is not free to choose but should be determined by the following manner. We fit the experimental data (i.e., the relative intensity explained above) using the function $c_{\mathrm{fit}} = c_0 \exp(\gamma \cos(\theta-\theta^\prime))+c^\prime$ to obtain $\gamma$, where $\theta^\prime$ and $c^\prime$ are auxiliary fitting parameters to compensate the unavoidable shifts in the experimental results. 
Here, $c_0$ is determined so that $\int_{-\pi}^\pi c_{\mathrm{fit}}\mathrm{d}\theta=1$, because the experimental data are normalized by the average over whole $\theta$ as described above. 
Thus, $c_0$ is a function of $\gamma$ and $c^\prime$. 
By noting the identity $\int_{-\pi}^\pi \exp(\gamma\cos\theta)\mathrm{d}\theta = 2 \pi \mathcal{I}_0(\gamma)$, where $\mathcal{I}_0$ is the modified Bessel function of the first kind, we get $c_0=(1-2\pi c^\prime)/(2\pi \mathcal{I}_0(\gamma))$. 
The function $\mathcal{I}_0$ can be approximated by an elementary function \cite{Olivares2018}, which is used in the actual fitting. 
The above-described procedure is carried out using the software ImageJ. 

The adequateness of the fitting is evaluated by the coefficient of determination $\mathsf{R}^2$, where $\mathsf{R}^2(<1)$ close to unity indicates the successful fitting.
The values of $\mathsf{R}^2$ for the fitting procedure are summarized in Table~\ref{tab:R2}. 
For the cases with the carboxylate-modified target particles, the signal-to-noise ratio is good and the model \eqref{eq:c} fits the experimental data well, i.e., $\mathsf{R}^2 \approx 1$. 
For the cases with the amine-modified target and the case without the target particle, the value of $\mathsf{R}^2$ is decreased to $>0.6$ for $P=78.7$~mW, but still the model \eqref{eq:c} can be considered to fit the experimental data. 
In the analysis of the present paper, the data with bad values of $\mathsf{R}^2<0.6$ due to stochastic noise are not used in the case without the target particle.

\begin{table*}[tb]
    \centering
        \caption{The coefficient of determination $\mathsf{R}^2$ for the fitting procedure used in the present paper. For typical examples of the actual fitting curves, see Figs.~\ref{fig:NoPS}, \ref{fig:main}, and \ref{fig:Main-NH2}. For the actual fitting values, see Fig.~\ref{fig:PREcompare}(a). 
        }
    \begin{tabular}{ccccccccc}
\hline
                & ~  & \multicolumn{3}{c}{carboxylate-modified}& ~ & amine-modified & ~ & no target particle \\
$R$~[\textmu m] & ~  & $P=78.7$~mW & $P=46.7$~mW & $P=30.2$~mW & ~ & $P=78.7$~mW    & ~ & $P=78.7$~mW        \\ \cline{3-5}\cline{7-7}\cline{9-9}
$4.4$           & ~  & $0.97$      & $0.98$      & $0.97$      & ~ & $0.61$         & ~ & $0.73$             \\ 
$5.4$           & ~  & $0.97$      & $0.98$      & $0.85$      & ~ & $0.25$         & ~ & ---                \\
$6.5$           & ~  & $0.94$      & $0.92$      & $0.84$      & ~ & $0.01$         & ~ & ---                \\
$7.5$           & ~  & $0.86$      & $0.69$      & $0.66$      & ~ & $0.01$         & ~ & ---                \\
\hline 
    \end{tabular}
    \label{tab:R2}
\end{table*}

\bibliography{00ref}

\end{document}